# About the description of physical reality of Bell's experiment.

## Alejandro A. Hnilo

*CEILAP, Centro de Investigaciones en Láseres y Aplicaciones, (MINDEF-CONICET);*
*J.B. de La Salle 4397, (1603) Villa Martelli, Argentina.*
*email: alex.hnilo@gmail.com*

*Abstract.*

A hidden variables model complying with the simplest form of Local Realism was recently introduced, which reproduces Quantum Mechanics' predictions for an even ideally perfect Bell's experiment. This is possible thanks to the use of a non-Boolean vector hidden variable. Yet, that model is as far as Quantum Mechanics from the goal of providing a complete description of physical reality in the EPR-sense. Such complete description includes the capacity to calculate, from the values taken by the hidden variables, the time values when particles are detected. This can be achieved by replacing Born's rule (which allow calculating only probabilities) with a deterministic condition for particle detection. The simplest choice is a threshold condition on the hidden variables. However, in order to test this choice, a new type of quantum (or wave, or non-Boolean) computer is necessary. This new type of quantum computer does not exist yet, not even in theory. In this paper, a classical (Boolean) computer code is presented which mimics the operation of that new type of quantum computer by using contextual instructions. These instructions take into account a consequence of the principle of superposition (which is a typical vector, i.e. non-Boolean, feature). Numerical results generated by the mimicking code are analyzed. They illustrate the features the hypothetical new type of quantum computer's output may have, and show how and why some intuitive assumptions about Bell's experiment fail.

*September 6th, 2021.*

# 1. Introduction.

Since the famous 1935 paper by Einstein, Podolsky and Rosen (EPR) [1], many hidden variables (HV) theories have been proposed to complete the description of physical reality provided by Quantum Mechanics (QM). In 1965, by analyzing an experiment similar to the one sketched in Figure 1, J.Bell formalized the notion that the completion was an impossible task without breaking intuitive ideas of Locality, or Realism [2] (in short: Local Realism). Recently [3], the subject was split in two: the problem of the apparent contradiction between QM and Local Realism was named the first or "soft" problem. The complete description of physical reality in the EPR-sense involves some mechanism able to predict the *time values* when particles are detected. The resulting rates of single and coincident detections must fit the QM predictions, even in ideal conditions. This has been named the second or "hard" problem, because its solution (assuming it is achieved within Local Realism) also solves the first one.

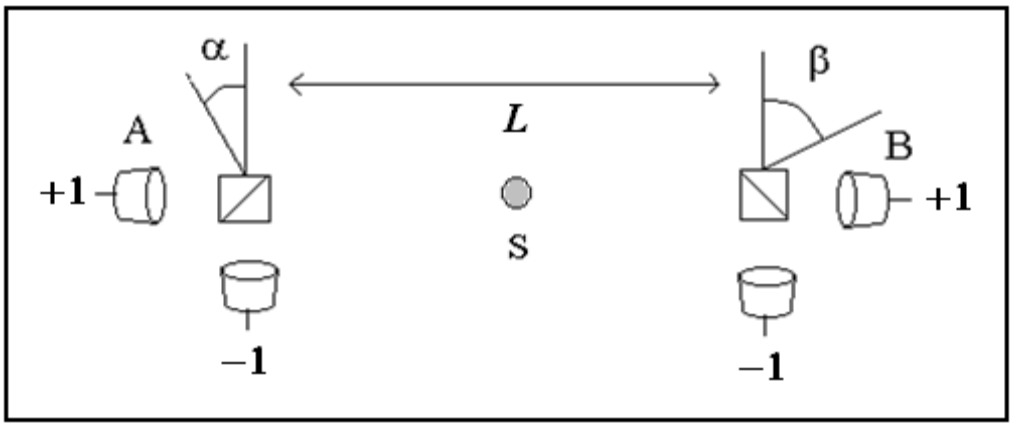

Figure 1: Sketch of a typical Bell's (or EPR-Böhm's) experiment. Source **S** emits pairs of systems entangled in polarization, which propagate to stations **A** and **B** separated by a (large) distance $L$, and are detected after being transmitted (+1) or reflected (-1) by polarization analyzers oriented at angles ($\alpha,\beta$). Simultaneous outcomes at each station are recorded; their correlation violates classical limits (Boole-Bell's inequalities).

The "soft" problem was solved in [3] by introducing a simple example, the so named *vector-HV* model, which demonstrates QM predictions for the Bell's experiment to be compatible with the simplest, non-Boolean form of Local Realism. That is, a definition of Local Realism that does not involve the use of classical probability. Classical probability holds to Kolomogorov's axioms, which in turn are based on Boolean logic. In fact, the cause of the apparent contradiction between QM and Local Realism is the use of classical probability in the derivation of Bell's inequalities. Bell's inequalities are equivalent to Boole's conditions of completeness [4,5]. Therefore, violating Bell's inequalities is a logical impossibility for any Boolean model. The vector-HV model is non-Boolean and, in spite of clearly holding to (non-Boolean, of course) Local Realism, reproduces QM predictions. This result should end decades of discussions.

Nevertheless, the situation is not fully satisfactory. The vector-HV model, as described in [3], is not able to predict the time values when particles are detected. It is not closer than QM to solve the "hard" problem. It fails to predict those time values because, in order to determine particle detection, it applies a threshold condition that is valid only in the statistical regime (i.e., many particles detected). The likely next step is to extend the vector-HV model and the threshold condition to the single particle regime. This approach is believed to solve the "hard" problem if run into a non-Boolean framework, i.e., a new type of quantum computer (NTQC). A NTQC should have the "wavy" nature of quantum computers [6], accept vector-HV values as input data, and give up Born's rule to link wave amplitude (or vector modulus) with particle detection [7]. But NTQC are not currently available, not even in theory.

At this point, a natural question is: isn't there a code running in a classical computer able to *mimic* the operation of a NTQC, to get an idea of how the actual solution may look like? The answer is "yes", but at the cost of inserting *contextual* instructions in the classical code. This alternative to reproduce QM results within a Boolean framework is not new.

This paper deals with the justification and realization, and the communication of the main results, of that mimicking classical code. In the next Section 2, the extension of the vector-HV threshold condition to the single particle regime is outlined. The reasons why contextual instructions are unavoidable in a Boolean code are reviewed. In the Section 3, the operation of the mimicking classical code is detailed, and some numerical results are analyzed. As all HV models, this one is not claimed to represent actual physical processes. Its aim is merely to show how the results of a NTQC and the solution of the "hard" problem may look like. It shows how and why some apparently natural assumptions about the Bell's experiment fail.

It is fair to say here that the solution of the "hard" problem is of academic interest only, for the input data (the values of the HV) necessary to calculate the time values of particle detection are, by definition, unknown. For all practical purposes, current quantum computers suffice.

## 2. The vector-HV model in the single particle regime.
*2.1 Non-Boolean logic.*

In the (non-Boolean) vector-HV model, features of physical systems are not represented by a set, but by a vector. Systems having more than one feature are not found by intersection of the sets representing these features, but by projection of vectors. In what follows, vectors are indicated with bold typing, scalars with normal typing. The operation *projection* of vector **b** into vector **a** is:

$$\mathbf{a}.\mathbf{b} \equiv |a.b.cos(\gamma)|.\mathbf{e}_a \qquad (1)$$

where $\gamma$ is the angle between **a** and **b**, and $\mathbf{e}_a$ is the unit vector in the direction of **a**. This operation is neither commutative nor associative, and implies a non-Boolean algebra. This single reason suffices for the vector-HV model to be able, in principle, to violate Boole-Bell's inequalities. Eq.1 looks like the projection of state $|b\rangle$ into state $|a\rangle$ in QM: $|a\rangle\langle a|b\rangle$. Yet, be aware that the elements of QM algebra are not simple vectors, but closed vector subspaces or their corresponding orthogonal manifolds in a Hilbert's space. QM algebra is an ortho-complemented non-distributive lattice [8], and is (of course) non-Boolean too.

Nevertheless, vectors explain in simple terms many quantum features: interference, superposition, and operators' non-commutativity. The main remaining problem is how to link vectors' length (or wave amplitude), which are continuous variables, to discontinuous particle detection. This has been identified as the true quantum problem [7]. Born's rule is the simplest solution: the squared vector's length is the probability of detecting a particle. Yet, by relating vector's modulus with *probability*, Born's rule confines QM to the statistical regime and gives up any hope of calculating the time value of particle detection. The simplest choice to link vectors' length to deterministic particle detection is to use some threshold condition. This was done in [3] but restricted to the statistical regime; here it is extended to the single particle one.

*2.2 The threshold condition.*

Following the approach in [3], assume that the system under consideration carries a vector HV, named $\mathbf{V}(t)$:

$$\mathbf{V}(t) = f(t).\mathbf{e}_x + g(t).\mathbf{e}_y = V(t).\mathbf{v}(t) \qquad (2)$$

where $\mathbf{e}_x$, $\mathbf{e}_y$ are unit vectors in the directions perpendicular to the system's propagation direction, and $f(t)$ and $g(t)$ are functions of time. The modulus of $\mathbf{V}(t)$ is $V(t)$, the unit vector in its direction is $\mathbf{v}(t)$, which is at an angle $\nu(t)$ with the *x*-axis. For a non-polarized ensemble of systems, $V(t)$ and $\nu(t)$ are statistically independent. The integral of the squared modulus of $\mathbf{V}(t)$ during time interval $[\theta_i, \theta_i+T]$ is named $m_i$:

$$m_i \equiv \int_{\theta_i}^{\theta_i + T} dt.|\mathbf{V}(t)|^2 \geq n.u \Rightarrow n \text{ particles are detected during } [\theta_i, \theta_i+T] \qquad (3)$$

or: $n$ = INTEGER ($m_i/u$), where $u$ is some threshold value. In [3], the validity of eq.3 was restricted to the statistical regime $n \gg 1$. Now this restriction is removed. If $n=1$, eq.3 allows determining the time interval $[\theta_i, \theta_i+T]$ in which one particle is detected. It must be kept in mind that $\mathbf{V}(t)$ is a hypothetical HV. It can have the features that are found convenient as far as no contradiction arises.

*2.3 Contextuality.*

If the vector-HV model is transcribed into a classical computer code with the threshold condition eq.3, the results fail to violate Bell's inequalities. This is not surprising, for Vorob'yev's theorem [9] demonstrates that violation of Bell's inequalities can occur within Boolean logic only in *contextual* scenarios. Some words on the meaning of contextuality are in order.

The elementary definition of contextuality is that a Boolean HV $\lambda$ has a probability distribution which is a function of the settings, say $\rho(\lambda,\alpha)$ in Fig.1. In a deeper approach [10], the space $\Omega$ of the HV is assumed a classical probability space. For, it is necessary to average over it in order to calculate observable *probabilities*. Contextuality then means that it is necessary to define a Boolean sub-algebra $\Omega_B$ for each physical quantity to be measured. These sub-algebras define the different contexts, or experimental setups.

In my opinion, contextuality is a meaningful term within Boolean logic only. If Boolean logic is given up (as it is done in the vector-HV model, or in QM), then no probability distribution, no classical probability space, no Boolean sub-algebra can be defined, and hence it is not possible to speak of contextuality or non-contextuality. When attempting to write down a Boolean computer code to describe Bell's experiment, contextual instructions immediately arise as necessary. Vorob'yev's theorem means that this necessity is the consequence of using Boolean logic. Non-Boolean (or quantum) computers do not need contextual instructions. Saying that the vector-HV model, or QM itself, is contextual is misleading. They are neither contextual nor non-contextual, they are non-Boolean.

Non-Booleanity implies that a description in terms of probabilities is not always possible. A suitable example is, precisely, the vector-HV. One may think that the angle variable $\nu(t)$ of a non-polarized **V**(t) has a uniform distribution, see Figure 2a. Yet, as a consequence of the principle of superposition (a non Boolean, vector feature) the distribution of $\nu(t)$ can also be thought as the sum of two Dirac's deltas centered at $\pi/2$ of each other (but that can be placed anywhere, see Fig.2b). Which of the distributions is the correct one? The uniform one or the two deltas? (and in the second case, which of the infinite possible ones?). In the intuitive (Boolean) way of thinking, only an infinitesimal fraction of the "uniformly distributed" values of $\nu(t)$ is parallel or orthogonal to a certain angle value $\alpha$. Yet, these fractions turn out to be here as large as ½, for *any* value of $\alpha$. This has the flavor of an inversion of cause and effect, a paradox often found in QM. But there is no such inversion. The correct answer is: no probability distribution is correct. The definition of (classical) probability requires Boolean logic, and vectors are non-Boolean. It is not possible to assign a probability distribution because of the (non-Boolean) superposition principle, i.e., the possibility to project **V**(t) into arbitrary axes.

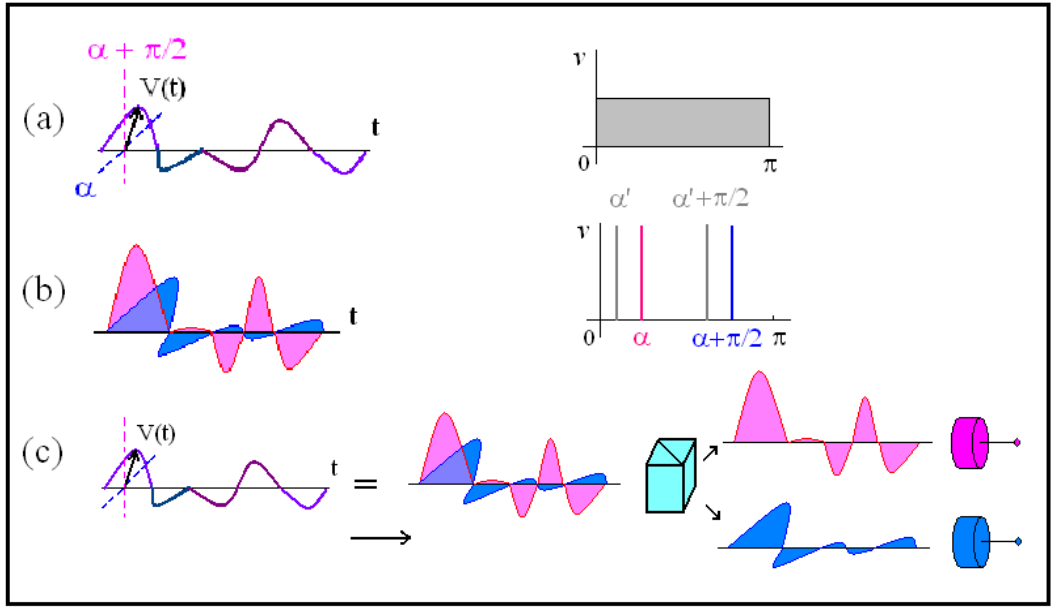

Figure 2: **(a)** *Left:* vector **V**(t) evolves inside and outside the plane of the paper (red in the plane, blue orthogonal to it, different degrees of violet indicate intermediate planes). *Right:* For a non-polarized ensemble, the "probability distribution" of ν(t) is intuitively thought uniform in [0,π]. **(b)**: *Left:* the vector can be projected into two components, one orthogonal to an arbitrary angle α (the plane of the paper) the other one parallel to it. *Right:* as a consequence, the "probability distribution" of ν(t) changes from uniform to two "deltas", placed at α and α+π/2. Besides, α' and α'+π/2 are also possible. This does not mean an inversion of cause and effect, but that a description in terms of probability distributions (which presupposes Boolean logic) is impossible. **(c)**: The vector components that determine whether one detection in each output gate occurs, or not, are well defined *before* the arrival to the analyzer. The analyzer merely separates two already existing components.

*2.4 Justification of the contextual instructions.*

The problem now is to find the form of the appropriate contextual instructions to mimic a NTQC in a classical computer. Consider the following idea: when a particle is detected at the +1(-1) gate of an analyzer, it is so because the component of **V**(t) parallel (orthogonal) to α passes (at that time value) the threshold condition for particle detection. This is obviously true after the analyzer; but it is also true *before* the analyzer. If this idea sounds strange, it is because one is used to think in QM terms. In QM, it is stressed that polarization is defined only after the analyzer. However, the idea should not sound strange at all, for a perfect analyzer does not interchange energy with the incident field. It does not modify nor scramble its components, it just splits them (Fig.2c). The idea simply means that the vector's component parallel (orthogonal) to α has reached the condition to produce detections, even before the analyzer. Be aware that no "quantum" effect is involved here, just the superposition principle and a perfect analyzer. The idea is applied here in relationship with the threshold condition eq.3, but it applies to any condition that determines detection, i.e., to any condition linking vector's size (or wave amplitude) with particle detection at a given time.

Let assume, for simplicity, that the modulus of **V**(t) is constant. Then the threshold condition involves only the angle ν(t). If a particle is detected after the analyzer in station A at time $θ_i$, it is because $v^A(\theta_i)$ has a certain relationship (= it is at a certain angle) with $\mathbf{e}_\alpha$. This feature is well defined even before arriving to the analyzer. According to the vector-HV model, in an entangled state $v^A(\theta_i)$ and $v^B(\theta_i)$ are related from the moment of emission (see [3]). Therefore, $v^B(\theta_i)$ has with $\mathbf{e}_\alpha$ the same relationship than $v^A(\theta_i)$ has. This is not action-at-a-distance, it is just the consequence of using HV to create correlations between remote places. In turn, the projection of $\mathbf{V}^B(t)$ into $\mathbf{e}_\beta$ determines whether a particle is detected at station B. In particular, the projection of $\mathbf{V}^B(\theta_i)$ (that is, the projection of the $\mathbf{V}^B$ vector at the time values when a particle is detected in station A) into $\mathbf{e}_\beta$ determines coincidences. Let call $T_\alpha$ ($T_{\alpha\perp}$) the set of time values when a particle is detected at the +1(-1) gate of an analyzer oriented at angle $\alpha$. Then the following statement is valid:

$$\forall \theta_i \in T_\alpha: \mathbf{v}(t)//\mathbf{e}_\alpha \text{ (or } |v(t)-\alpha|= 0,\pi) \text{ and } \forall \theta_i \in T_{\alpha\perp}: \mathbf{v}(t)\perp\mathbf{e}_\alpha \text{ (or } |v(t)-\alpha|= \pi/2) \qquad (4)$$

This is valid (for vectors) because of the superposition principle. A NTQC would have it embedded in its wavy (non-Boolean) nature. In order to mimic this feature in a Boolean computer, eq.4 must be "artificially" inserted in the code. In other words: each time $\theta_i$ a particle is detected in the +1(-1) gate of the analyzer in station A, then $v(\theta_i)$ is imposed to be parallel (orthogonal) to $\mathbf{e}_\alpha$. This illustrates the meaning of the often stated quantum principle: *the properties of a physical system depend on the definition of the apparatus of observation* (in this case, the value of $\alpha$). If a description within Boolean logic, or with a classical computer is attempted, then it is necessary to define the apparatus, i.e. the value of $\alpha$, i.e. the context.

The application of eqs.3-4 define a classical code (detailed in the Appendix) that reproduces statistical QM predictions exactly and besides, is able to predict *when* single detections and coincidences occur. It shows how the results of the completion of physical reality (of Bell's experiment) in the EPR-sense may look like. The code assumes that in each measurable time interval [$\theta_i$, $\theta_i+T$] one system is incident on each analyzer in Fig.1, that is, $m^A_i = m^B_i = u$. Each pair carries random angular variables $v^A_i$ and $v^B_i$, which are related in a different way depending on the Bell state emitted by the source. The contribution of $\mathbf{V}(t)$ to the +1 gate in station A is:

$$m^\alpha \equiv \int_0^{Tr} dt. |\mathbf{e}_\alpha \cdot \mathbf{V}(t)|^2 = \int_0^{Tr} dt. |cos(v^A(t) - \alpha).V(t)|^2 \qquad (5)$$

and in the same way $m^{\alpha\perp}$ for the -1 gate. Each time $\theta^A_i$ that $m^\alpha$ ($m^{\alpha\perp}$) cross an integer number of $u$, then one particle is detected at the +1(-1) gate. According to eq.4, $v^A_i = \alpha$ ($v^A_i = \alpha+\pi/2$) if the detection is observed at the +1 (-1) gate. In the case of the fully symmetrical Bell state $|\varphi^+_{AB}\rangle = (1/\sqrt{2}).\{|x_A,x_B\rangle+|y_A,y_B\rangle\}$, $v^A_i = v^B_i$ [3], hence, also $v^B_i = \alpha$ (or $\alpha+\pi/2$) at time $\theta_i$. Unless no particle at all is detected in station A, this value (i.e. $\alpha$ or $\alpha+\pi/2$, not the initial random value) is used in the analogous of eq.5 for B station to calculate $m^\beta$. Each time $\theta^B_i$ that $m^\beta$ ($m^{\beta\perp}$) reaches an integer multiple of $u$, one particle is detected at the +1(-1) gate in station B. Coincidences occur when $\theta^A_i = \theta^B_i$. This code is not the proper solution of the second (or "hard") problem because contextual instructions are inserted, but it illustrates how that solution may look like once NTQC are available.

## 3. Numerical results.

In the setup of Fig.1, $N_A^+$ is the number of detections in gate +1 in station A, and $N_{AB}^{++}$ is the number of detections +1 observed simultaneously in stations A and B, both recorded during a complete experimental run. In the same way: $N_A^-$, $N_{AB}^{+-}$, $N_B^+$, etc. The most important result is the curve of (f.ex.) $N_{AB}^{++}/N_A^+$ as a function of the difference angle $\alpha-\beta$ (for the Bell state $|\varphi^+_{AB}\rangle$), see Figure 3. It perfectly fits the QM prediction, $\frac{1}{2}\cos^2(\alpha-\beta)$.

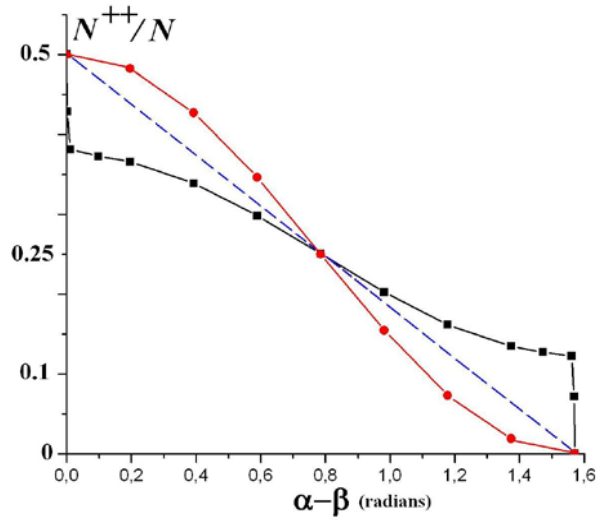

Figure 3: Rate of (+1,+1) coincidences as a function of the difference angle (in radians, for state $|\varphi^+_{AB}\rangle$) calculated ($N=10^6$) according to: vector-HV model with contextual instructions (red circles, this is also the QM prediction); the same, without contextual instructions (black squares). For $N\to\infty$, the latter coincides with a semi-classical theory excepting in $\alpha-\beta = 0$ and $\pi/2$, where it coincides with QM (see text). The dashed blue straight line is the correlation limit imposed by Boole-Bell's inequalities.

The next most important result is efficiency, which is defined for each gate. F.ex., the efficiency at gate +1 of station A is:

$$\eta^+_A = (N_{AB}^{++} + N_{AB}^{+-})/ N_A^+ \qquad (6)$$

In the ideal case, all the four efficiencies: $\eta^+_A$, $\eta^-_A$, $\eta^+_B$, $\eta^-_B$, are equal to 1. They provide a decisive evaluation of the quality and significance of a model or experiment, because classical HV models with efficiencies as high as (at least) $1/\sqrt{2} \approx 0.71$ are well known to exist. As shown in the Appendix, the code produces results with all efficiencies exactly equal to 1. Besides, the total number of detected particles after the analyzer (that is, summing up both gates) is equal to the number of particles that would-be detected if a detector were placed before it, minus *one*. The difference is irrelevant statistically and, of course, in the experiments.

The code allows exploring counterfactual situations, something that is impossible to do in the real world. For the *same* series of (random) values $\nu_i$ as input (this cannot be done in the real world), particles are detected in the +1 or -1 gates at different time values depending of the analyzer's orientation. F.ex., for time values between 20 and 40 and $\alpha=0$, in a given run particles are detected in station A at times:

+1 gate: 21,22,24,27,28,31,35,37,39,40.

-1 gate: 20,23,25,26,29,30,32,33,34,36.

If $\alpha=\pi/8$ instead:

+1gate: 20,21,22,24,26,27,28,31,33,35.

-1gate: 23,25,29,30,32,34,37,38,40.

If $\alpha=\pi/2$, time values that appeared in line +1 when $\alpha=0$ appear in line -1 instead, and vice versa (this explains the lack of drift between the A and B integrals if $\alpha-\beta = \pi/2$ that is mentioned below). Note that there is one particle detected in all time intervals, what fits the flux incident on the analyzer. The code's logic makes impossible that particles are simultaneously detected at both gates in the same station.

Meanwhile, in station B, the time values particles are detected at each gate if $\alpha=0$, $\beta=\pi/8$ (always for the same set of values of $\nu_i$ as input) are:

+1 gate in B: 21,23,26,28,32,34,36,38,40.

-1 gate in B: 20,22,24,25,27,29,30,31,33,35,37,39.

if $\alpha=\pi/4$, $\beta=\pi/8$ instead:

+1 gate in B: 21,22,24,25,28,29,30,34,36,38,39,40.

-1 gate in B: 20,23,26,27,31,32,33,35,37.

The distribution of time values of detection in B have changed, although it is the setting in station A what has changed (be aware that $\beta=\pi/8$ in both cases). In [11], it was assumed that the time series in one of the stations is independent of the settings in the other. Using only this "strong" counterfactual assumption, Bell's inequalities are derived without any reference to Local Realism. The mentioned result shows how and why that assumption fails in an "actual" Bell's experiment. Note that it fails at the observable level, *even if* it is strictly true at the HV level. In other words: the series of $\nu_i$ is always the same and is independent of the settings, but the series of observable coincidences is not. I stress this analysis is possible because we can run the code with the *same* values of $\nu_i$ for different settings. It would be impossible in an actual experiment, where the values of the HV are incontrollable.

Now, consider the curve:

$$N^{++} / N = ¼.[ ½ + cos^2(\alpha-\beta)] \tag{7}$$

which corresponds to a semi-classical radiation theory (SCRT). In SCRT, the polarization of the field is well defined at the moment of emission, varies randomly in time, and the probability of detecting a particle is proportional to the square of the field's amplitude. In other words, eq.7 is the result obtained if Born's rule is applied *separately* to correlated classical waves in each station (instead of applied to an object existing in an abstract four dimension space, as the quantum state $|\varphi^+_{AB}\rangle$ is). Of course, eq.7 does not violate Bell's inequalities. If contextuality is removed from the presented code (simply delete instructions *nu = alfa* and *nu = alfa + π/2* in code lines 105 and 205, then $v^B_i = v^A_i \forall i$), then the numerical results fit eq.7 excepting for the isolated points $\alpha-\beta = 0$ and $\alpha-\beta = \pi/2$, see Fig.3, curve in black.

It is remarkable that the number of detections at the +1(-1) gate of the analyzer in station A is proportional to $cos^2(\alpha-v_i)$ ($sin^2(\alpha-v_i)$), where $v_i$ is the value of the HV at the time $\theta^A_i$ the detection is observed. Taking into account that the HV is a vector, this means that the *average* orientation of the HV is parallel (perpendicular) to the analyzer's axis when a particle is detected at the +1(-1) gate. This can be clearly seen in a plot in polar coordinates, Figure 4. If these values of $v_i$ are directly used in station B, the SCRT results are obtained (excepting for $\alpha-\beta = 0, \pi/2$). If they are replaced by $v_i = \alpha$ or $\alpha+\pi/2$ instead (i.e., the contextual instructions), the QM results are obtained. Therefore, from the numerical point of view, the contextual instructions mean replacing each single $v_i$ by their average value $\langle v \rangle$.

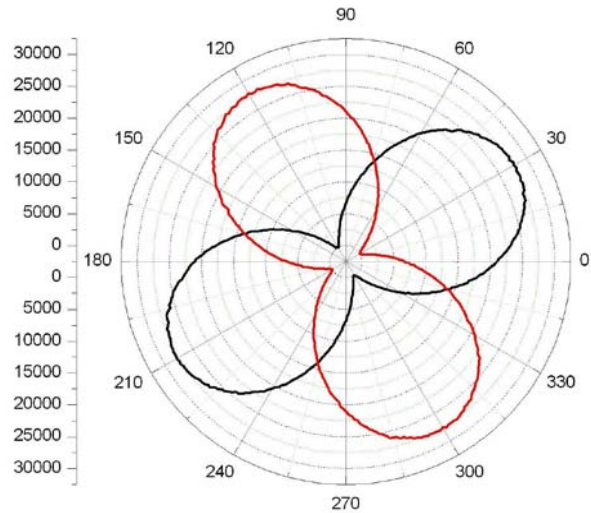

Figure 4: Polar plot of the number of detections as a function of $v$ at the time value a particle is detected in the +1 (black) and -1 (red) gates in station A, they fit the curves $cos^2(v-\alpha)$ and $sin^2(v-\alpha)$ respectively, $\alpha = 30°$, $N = 10^7$, resolution $1°$.

The numerical reason of the failure (to reproduce QM predictions) of the non-contextual version of the code is that the growth of the time integrals in A and B drift out of pace as time goes on. After a few thousands of time steps the integrals' growths become fully uncorrelated. In consequence, integer multiples of $u$ are reached simultaneously in both stations (what means one detected coincidence) only by chance, as if applying Born's rule separately to the vectors in each station. This is the reason why the SCRT curve is obtained for almost all angle settings. In the isolated points $\alpha-\beta = 0, \pi/2$ the integrals' growths remain synchronized during the whole run. In the first case, because the calculations in both stations are identical; in the second case, because they correspond to orthogonal gates in the same station (see some lines above).

An alternative to avoid the drift is to reset the integrals' values to zero each time a particle is detected. But, this has the undesirable consequence that efficiency drops to 50%. This is because "fractions of particle" are lost when the integrals are reset. In the average, half the incident particles are lost. This alternative is hence of little interest. Yet, it is noteworthy that the "reset" instruction does increase the level of correlation between stations in all cases. If it is applied to the non-contextual code, the correlation increases from the SCRT curve (and $S_{CHSH} = \sqrt{2}$) up to the (blue, dashed in Fig.3) straight line that is the limit allowed by Bell's inequalities (and $S_{CHSH} = 2$). If it is applied to the contextual code instead, the correlation increases from the QM curve and $S_{CHSH} = 2\sqrt{2}$ up to a super-quantum correlation value $S_{CHSH} = 3$. In all cases, efficiencies remain near 50%.

The code displayed in the Appendix is set to the rate of one pair detected per unit time before the analyzers. This setting produces perfect efficiencies and entanglement. A higher rate (say, Vmed=2, $u$=1, or two particles detected per unit time in the average) deteriorates the level of entanglement (the value of $S_{CHSH}$), as it occurs in the experiments. A lower value (say, Vmed=1, $u$=10, or one particle detected each ten time units in the average) reduces efficiency as $\approx 1/u$.

It has been noted that the way this code operates follows Lüders postulate [12,13] and is hence more *nonlocal* than *contextual*. In fact, only the distribution of the HV for station B depends on the settings. In other words: it is $\rho(\alpha,\lambda)$ for station B, while for station A is just $\rho(\lambda)$. Of course, if a distribution $\rho(\alpha,\lambda)$ or $\rho(\alpha,\beta,\lambda)$ is used in both stations, the QM predictions are reproduced too. Discussing whether the code, as it is written here, deserves to be qualified as "contextual", "half-contextual", "Lüder postulate based" or "nonlocal" is an issue I find of little interest. For, the code is just an example of how to mimic, with a Boolean code, the vector (i.e., non-Boolean) feature illustrated in Fig.2. Far more interesting, in my opinion, is to put effort in imagining some design for a NTQC.

**Summary.**

The "EPR paradox" has been split in two chained problems: the contradiction between QM and Local Realism is named the first or "soft" problem. The second or "hard" problem is to complete the description of physical reality in the EPR-sense, and means predicting not only detection rates, but also the time values when particles are detected. Solving the second problem also solves the first one. The first problem has been properly solved by using a vector (non-Boolean) HV and a threshold condition for particle detection valid only in the statistical ($N$>>1) limit. Yet, that solution (detailed in [3]) is no closer than QM to solve the second or "hard" problem.

In order to solve the second problem, it is necessary a computer code that, having **V**(t) as input data, prints the time value of single (and in consequence, also coincident) detections. But, proper calculations of non-Boolean problems with classical (Boolean) computer codes are impossible. Quantum computers are non-Boolean indeed but, in their current form, they use Born's rule and do not accept HV as input data. Quantum computers using eq.3 instead of Born's rule are named here NTQC and have not been devised yet. In the meantime, what is at hand is to *mimic* the operation of a non-Boolean computer, by artificially introducing a vectors' feature (which is a direct consequence of the principle of superposition) into the Boolean code. This is in agreement with Vorob'yev's theorem. The obtained numerical results show some features the NTQC's output may have. Some features cannot be experimentally observable, in particular: the way contextuality manifests itself in the data, and why "strong" counterfactual assumptions fail.

Finally, the idea that a NTQC can solve the second problem may seem in contradiction with the demonstration that, assuming that free will exists, no extension of QM (roughly speaking: a HV theory) can provide more information (about the outcomes of future measurements) than QM itself [14]. That demonstration is derived from constraints on distributions of conditional probabilities; hence, it presupposes Boolean logic. As neither the underlying vector-HV model nor the hypothesized NTQC operate according to Boolean logic, there is no contradiction.

**Acknowledgements.**
Many thanks to Mónica Agüero, Donald Graft, Federico Holik and Marcelo Kovalsky for the interest and advices. This material is based upon research supported by, or in part by, the U.S. Office of Naval Research under award number N62909-18-1-2021. Also, by grant PIP 2017-027 CONICET (Argentina).

**References.**
[1] A.Einstein, B.Podolsky and N.Rosen, " Can quantum-mechanical description of physical reality be considered complete?", *Phys.Rev.* **47**, p.777 (1935).
[2] J.Clauser and A.Shimony, "Bell's theorem: experimental tests and implications", *Rep. Prog. Phys.* **41** p.1881 (1978).
[3] A.Hnilo, "Non-Boolean Hidden Variables model reproduces Quantum Mechanics' predictions for Bell's experiment", *arXiv/quant-ph/*:2005.10367
[4] H. De Raedt *et al.*, "Extended Boole-Bell Inequalities Applicable to Quantum Theory", *J. of Computational and Theoretical Nanoscience*, **8** p.1011 (2011).

[5] I.Pitowski, "George Boole's 'Conditions of Possible Experience' and the Quantum Puzzle", *Brit. J. Phil. Sci.* **45** p.95 (1994).

[6] N.David Mermin, "What has Quantum Mechanics to do with factoring?" *Phys.Today* **60**(4) p.8 (2007).

[7] A.Khrennikov, "Quantum probabilities and violation of CHSH-inequality from classical random signals and threshold type detection scheme", *Prog. Theor. Phys.* **128**, p.31 (2012).

[8] S.Fortín *et al.*, "A logical approach to the Quantum-Classical transition", in *Quantum Worlds: Perspectives in the ontology of Quantum Mechanics*, O.Lombardi, S.Fortín, C.López and F.Holik Editors, Cambridge University Press, Cambridge 2019.

[9] N.Vorob'yev, "Markov measures and Markov extension", *Theory Prob. Appl.* **8** p.420 (1963) as quoted by A.Cabello, in "Historical introduction to quantum contextuality", in *Contextuality in Quantum Mechanics and beyond*, October 23rd 2020, https://www.youtube.com/watch?v=5r6J-PQVo_s.

[10] E.Beltrametti and G.Cassinelli, *The logic of Quantum Mechanics* Chapter 15, 25 and 26, Addison-Wesley, Reading 1981.

[11] L.Sica, "Bell's inequalities I: an explanation for their experimental violation", *Opt. Commun.* **170** p.55 (1999).

[12] M.Kupczynski, "Is the Moon There If Nobody Looks: Bell Inequalities and Physical Reality", *Frontiers in Physics* 8:273 (2020), doi: 10.3389/fphy.2020.00273.

[13] A.Khrennikov, "Two Faced Janus of Quantum Nonlocality", *Entropy* **2020**, 22, 303; doi:10.3390/e22030303

[14] R.Colbeck and R.Renner, "No extension of quantum theory can have improved predictive power", *Nat. Comm.* **2**:411 (2011).

## APPENDIX. A classical and contextual code.

This code follows the vector-HV model and the threshold condition in the single-particle regime in a straightforward way. It uses contextual instructions, what is unavoidable to mimic non-Boolean problems with a Boolean computer (for Vorob'yev's theorem). It is written for the Bell state $|\varphi^+_{AB}\rangle$ = $(1/\sqrt{2}) \cdot \{|x_A, x_B\rangle + |y_A, y_B\rangle\}$ and is asymmetrical in stations A and B. A symmetric version can be easily written, it is just twice longer. The code reproduces QM predictions *and besides*, it predicts *when* one particle is detected and one coincidence (of each type) occurs. Language is QBASIC, which can be freely downloaded from many websites. Comments are on the code lines:

```
INPUT "alfa, beta = "; alfa, beta      !input: α,β in radians.
INPUT "time = "; N                     !input: number of time intervals.
Vmed = 1: u = 1                        !single detection/interval condition
FOR i = 1 TO N                         !loop starts (time runs)
AT = 0: AR = 0: BT = 0: BR = 0         !reset flags of particle detection
nu = 2 * π * RND                       !random choosing ν
ka = (COS(alfa - nu)) ^ 2              !calculus of cos²(α-ν)
malfa = malfa + ka * Vmed              !integrate in gate A transmitted (T).
malfap = malfap + (1 - ka) * Vmed      !integrate in gate A reflected (R).
IF malfap > malfa THEN GOTO 90         !check if detect in T or in R.
IF malfa > u THEN GOTO 105             !check if detect in T.
GOTO 300
90 IF malfap > u THEN GOTO 205         !check if detect in R.
GOTO 300
105 nu = alfa                          !CONTEXTUAL INSTRUCTION
malfa = malfa – u                      !decrease integral in T.
NAtrns = NAtrns + 1: AT = 1            !Sum up detection, set flag A Transm.
GOTO 300
205 nu = alfa + π/2                    !CONTEXTUAL INSTRUCTION
malfap = malfap – u                    !decrease integral in R.
NArfl = NArfl + 1: AR = 1              !sum up detection, set flag A Refl.
300                                    !calculi in station A end.
kt = (COS(beta - nu)) ^ 2              !calculus of cos²(β-ν)
mbeta = mbeta + Vmed * kt              !integrate in gate T.
mbetap = mbetap + Vmed * (1 - kt)      !integrate in gate R.
IF mbetap > mbeta THEN GOTO 92         !check if detect in T or R.
IF mbeta > u THEN GOTO 405             !check if detect in T.
GOTO 600
92 IF mbetap > u THEN GOTO 505         !check if detect in R.
GOTO 600
405 mbeta = mbeta – u                  !decrease integral,
NBtrns = NBtrns + 1: BT = 1            !sum up detection, set flag B Transm.
GOTO 600
505 mbetap = mbetap – u                !decrease integral,
NBrfl = NBrfl + 1: BR = 1              !sum up detection, set flag B Refl.
600                                    !calculi in station B end.
IF AT * BT = 1 THEN NTT = NTT + 1      !sum up coincidences N⁺⁺
IF AT * BR = 1 THEN NTR = NTR + 1      !sum up coincidences N⁺⁻
IF AR * BT = 1 THEN NRT = NRT + 1      !sum up coincidences N⁻⁺
IF AR * BR = 1 THEN NRR = NRR + 1      !sum up coincidences N⁻⁻
NEXT i                                 !loop ends (end of run)
PRINT "NAtrans, NArefl= "; NAtrns, NArfl
PRINT "NBtrans, NBrefl= "; NBtrns, NBrfl
PRINT "NTT, NTR = "; NTT, NTR
PRINT "NRT, NRR = "; NRT, NRR
END
```

The Reader interested in reproducing results is advised to start by checking the following run:

Run program, and:

$$0, 0.3927 \quad \text{<enter>}$$
$$100000 \quad \text{<enter>}$$

You should get (warning: small variations may occur due to the RND function)

NAtrns = 49999, NArefl= 50000

NBtrns = 49999, NBrefl= 50000

$N^{++}$= 42760, $N^{+-}$ = 7239

$N^{-+}$= 7239, $N^{--}$ = 42761

Comparing the number of particles detected after the analyzers with the number that *would-be* detected if the analyzers *were not* in place, one particle is missing regardless the total number of particles. The difference is irrelevant in the statistical limit. The "missing" particle is hidden in the fractional values remaining in the integral variables, namely: *malfa, malfap, mbeta, mbetap* at the end of the run (these values are always smaller than 1 at the end of each loop). In the statistical limit, no particle is lost.

Instead, efficiencies as they are usually defined are *strictly* perfect, f.ex:

$\eta^+_A = (N^{++} + N^{+-})/ $ NAtrns= (42760+7239)/49999 = 1

After running the code for the usual angle settings and time $N=10^5$, $S_{CHSH} = 2.837 \approx 2\sqrt{2} > 2$.